\documentclass[preprint, review,12pt]{elsarticle}
\usepackage{amssymb}
\usepackage{amsmath}

\journal{Mathematical Biosciences}

\begin{document}

\begin{frontmatter}



\title{Travelling wave analysis of a mathematical model of glioblastoma growth}


\author{Philip Gerlee$^{1,\star}$}

\author{Sven Nelander$^{2}$}
\cortext[cor1]{Corresponding author: philipgerlee@gmail.com}
\address{$^1$ Mathematical Sciences, Chalmers University of Technology and G\"oteborg University, Chalmers Tv\"argata, 412 96 G\"oteborg, Sweden
\\$^2$ Department of Immunology, Genetics and Pathology, Rudbeck Laboratory, Uppsala University, 751 85 Uppsala, Sweden.}
\begin{abstract}
In this paper we analyse a previously proposed cell-based model of glioblastoma (brain tumour) growth, which is based on the assumption that the cancer cells switch phenotypes between a proliferative and motile state (Gerlee and Nelander, PLoS Comp. Bio., 8(6) 2012). The dynamics of this model can be described by a system of partial differential equations, which exhibits travelling wave solutions whose wave speed depends crucially on the rates of phenotypic switching. We show that under certain conditions on the model parameters, a closed form expression of the wave speed can be obtained, and using singular perturbation methods we also derive an approximate expression of the wave front shape. These new analytical results agree with simulations of the cell-based model, and importantly show that the inverse relationship between wave front steepness and speed observed for the Fisher equation no longer holds when phenotypic switching is considered. 


\end{abstract}


\end{frontmatter}


\clearpage
\section{Introduction}
The brain tumour glioblastoma kills approximately 80 000 people per year worldwide, and these patients have, despite decades of intense research, a dismal prognosis of approximately 12 months survival from diagnosis. The standard treatment is surgery, followed by radiotherapy and chemotherapy. However, one of the major hurdles in treating malignant glioblastomas surgically is their diffuse morphology and lack of distinct tumour margin. The high migration rate of glioblastoma cells is believed to be a main driver of progression \cite{Giese2003}, but precise knowledge of how glioblastoma growth is shaped by the underlying cellular processes, including cell migration, proliferation and adhesion, is still lacking, hampering the prospects of novel therapies and drugs.

One characteristic of glioblastoma cells which has gained  considerable attention is the `go or grow'--hypothesis, which states that proliferation and migration are mutually exclusive phenotypes of glioblastoma cells \cite{Giese2003}. This observation was recently confirmed using single cell tracking \cite{Farin2006}, where individual cells were observed to switch between proliferative and migratory behaviour. In order to understand and control the growth of glioblastomas we hence need an appreciation of how the process of phenotypic switching influences glioblastoma growth and invasion. This paper presents a starting point for this understanding and reports on an analytical connection between cell-scale parameters and the properties of tumour invasion, which could be used for tailoring treatment based on single-cell measurements.  



\section{Previous work}
The starting point of glioblastoma modelling was the seminal work of Murray and colleagues \cite{Tracqui1995,Woodward1996}, which made use of the Fisher equation
\begin{equation}\label{eq:fisher}
\frac{\partial u}{\partial t}  = D \frac{\partial^2u}{\partial x^2} + \rho u(1-u)
\end{equation}
where $u(x,t)$ denotes the density or concentration of cancer cells, $D$ is the diffusion coefficient of the cells, and $\rho$ is the growth rate. The microscopic process that the above equation describes is that of cells moving according to a random walk, and simultaneously dividing at rate $\rho$. It can be shown that the Fisher equation exhibits travelling wave solutions, which medically corresponds to a tumour invading the healthy tissue. These solutions $U(z)$ remain stationary in a moving frame with coordinates $z=x-ct$, and it can be shown that velocity of the invading front is given by $c=2\sqrt{D \rho}$.

Since then many different models of glioblastoma growth have been proposed, ranging from game theoretical models \cite{Basanta2011}, and systems of partial differential equations \cite{Swanson2011}, to individual-based models \cite{Khain2012}. In particular there has been an interest among modellers in the above mentioned 'go-or-grow' hypothesis, and several different approaches have been utilised. Hatzikirou et al. \cite{Hatzikirou2010} used a lattice-gas cellular automaton in order to investigate the impact of the switching between proliferative and migratory behaviour, and went on to show that in the corresponding macroscopic (Fisher) equation, there is a tradeoff between diffusion and proliferation reflecting the inability of cells to migrate and proliferate simultaneously. Similar results where obtained by Fedotov and Iomin \cite{Fedotov2008} but with a different type of model known as continuous time random walk model, where the movement of the cells is not constrained by a lattice. The effects of density-driven switching were investigated with a two-component reaction diffusion system in a study by Pham et al.\ \cite{Pham2011}, and they could show that this switching mechanism can produce complex dynamics growth patterns usually associated with tumour invasion.

In this paper we will be concerned with the analysis of an individual-based model put forward by Gerlee and Nelander \cite{Gerlee2012}. In the initial study, it was shown that the average behaviour of the cell-based model can be described by a set of coupled PDEs, similar to the Fisher equation, which exhibit travelling wave solutions. A combination of analytical and numerical techniques made it possible to calculate the wave speed of the solutions, and it was shown to closely approximate the velocity of the tumour margin in the cell-based model.

In this paper we extend the analysis of the model, and show that if one assumes that cell migration occurs much faster than proliferation, then a closed form expression of the wave speed can be obtained, and also that an approximate solution for the front shape can be derived. The paper is organised as follows: in section 3 we present the cell-based model and its continuum counter-part. Section 4 is concerned with obtaining a closed form expression for the wave speed, and in section 5 we derive an asymptotic solution to the system. Finally we conclude and discuss the implications of the results in section 6.


\section{The model}
The cells are assumed to occupy a $d$-dimensional square lattice containing $N^d$ lattice sites, and each lattice site either is empty or holds a single glioma cell. For the sake of simplicity we do not consider any interactions between the cancer cells (adhesion or repulsion), although this could be included \cite{Deroulers2009}.

The behaviour of each cell is modelled as a time continuous Markov process where each transition or action occurs with a certain rate, which only depends on the current and not previous states. Each cell is assumed to be in either of two states: proliferating or migrating, and switching between the states occurs at rates $q_p$ (into the P-state) and $q_m$ (into the M-state). A proliferating cell is stationary, passes through the cell cycle, and thus divides at a rate $\alpha$. {The daughter cell is placed in one of the empty neighbouring lattice sites (using a von Neumann neighbourhood) with uniform probability across all empty neighbouring sites.} If the cell has no empty neighbours cell division fails. A migrating cell performs a size exclusion random walk, where each jump occurs with rate $\nu$ (with dimension s$^{-1}$). {When motion is initiated the cell moves into one of the empty neighbouring lattice sites with uniform probability across all empty neighbouring sites. If the cell has no empty neighbours cell migration fails.}


{}

Lastly, cells are assumed to die, through apoptosis, at a rate $\mu$ (with dimension s$^{-1}$) independent of the cell state. This model is naturally a gross simplification of the true process of glioblastoma growth, and for further discussion on this we refer the reader to \cite{Gerlee2012}.

The time scale is chosen such that $\alpha = 1$, which means that all other rates are given in the unit 'cell cycle$^{-1}$'. Experimental results suggest that the average time for the cell cycle is 16-24 hours \cite{Giese2003}, and that the phenotypic switching occurs on a faster time scale than cell division \cite{Farin2006}, roughly on the order of hours, implying that $q_{p,m} \in (10,30)$. The death rate for an untreated tumour is on the other hand much smaller than the proliferation rate, approximately $\mu \sim 10^{-1}-10^{-2}$. Tracking of single cells has shown that glioblastoma cells move with a velocity of up to 25 cell sizes/cell cycle \cite{Farin2006}, and consequently we set $\nu = 25$.

\begin{figure}[!hptb]
\begin{center}
\includegraphics[width=10cm]{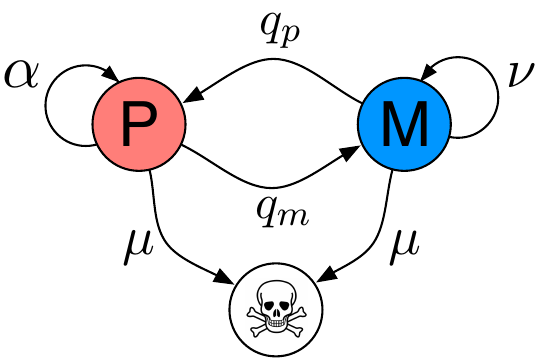}
\caption{\label{fig:model}A schematic of the continuous-time Markov chain which controls the behaviour of each cell in the individual-based model. The cells are either in a proliferative state (P) in which they divide at rate $\alpha$ or in a motile state where they jump between lattice points at rate $\nu$. The switching between the two states occurs at rate $q_p$ and $q_m$.} 
\end{center}
\end{figure}

The stochastic process behind the phenotypic switching is depicted schematically in figure \ref{fig:model}A. When comparing the cell-based model with the analytical results we simulate the model in $d=1$ dimensions. Each simulation is started with a single cell in the proliferative state at grid point $i=0$. We record the cell density at $t=T_{\max}/2$ and $t=T_{\max}$, and by performing a large number of simulations we estimate the occupation probabilities $\mathcal{P}_{i}^t$ and $\mathcal{M}_{i}^t$ of having a proliferating/migratory cell at lattice site $i$ at time $t$. By finding the lattice point where $\mathcal{P}_{i}^t + \mathcal{M}_{i}^t=1/2$ for $t=T_{\max}/2$ and $T_{\max}$ we can calculate speed of the advancing front. {If several such lattice points exist we pick the one with the smallest $i$.} Typically the probabilities are estimated from 20 different simulations and $T_{\max} = 100$ cell cycles.



\subsection{The continuum approximation}
The system of PDEs that describes the average behaviour of the cell-based model in one dimension was derived in Gerlee and Nelander \cite{Gerlee2012} and is given by:
\begin{align} \label{eq:origpde1}
&\frac{\partial p}{\partial t}  =D_{\alpha}(1-p-m)\frac{\partial^2p}{\partial x^2} + \alpha p (1-p-m) - (q_m + \mu)p + q_pm\\
&\frac{\partial m}{\partial t}  = D_{\nu} ((1-p) \frac{\partial^2m}{\partial x^2} + m \frac{\partial^2p}{\partial x^2})  - (q_p + \mu)m + q_mp \label{eq:origpde2}
\end{align}
where $p(x,t)$ and $m(x,t)$ is the density of proliferating and motile cells respectively. The diffusion coefficient $D_\alpha = \alpha/2$ captures tumour expansion driven by proliferation, while $D_\nu = \nu/2$ comes from the random movement of migratory cells.  The wave speed of this system can be determined by numerical investigation of the corresponding 4-dimensional autonomous system (for details see \cite{Gerlee2012}). Here we show how the system can be simplified and the problem reduced to three dimensions, which allows for a closed form expression of the wave speed. 


\subsection{Simplified system}
From the above parameter estimation we know that $\alpha = 1$ (due to the time scale chosen) and $\nu \approx 25$. This means that $\alpha \ll \nu $, and further implies that $D_\alpha \ll D_\nu$, which allows for a simplification of the system. We introduce the rescaling $\tilde{x}=x/\sqrt{D_\nu}$, which transforms \eqref{eq:origpde1}-\eqref{eq:origpde2} to
\begin{align} \nonumber
&\frac{\partial p}{\partial t}  = \frac{D_\alpha}{D_\nu}(1-p-m)\frac{\partial^2p}{\partial \tilde{x}^2} + \alpha p (1-p-m) - (q_m + \mu)p + q_pm\\
&\frac{\partial m}{\partial t}  = (1-p) \frac{\partial^2m}{\partial \tilde{x}^2} + m \frac{\partial^2p}{\partial \tilde{x}^2}  - (q_p + \mu)m + q_mp \nonumber
\end{align}
where $D_\alpha/D_\nu \ll 1$. Consequently we drop the diffusion term from the first equation, but return to the original space variable $x$ and end up with the following system:
\begin{align} \label{eq:pde1}
&\frac{\partial p}{\partial t}  =  \alpha p (1-p-m) - (q_m + \mu)p + q_pm\\
&\frac{\partial m}{\partial t}  = D_{\nu} \left((1-p) \frac{\partial^2m}{\partial x^2} + m \frac{\partial^2p}{\partial x^2} \right) - (q_p + \mu)m + q_mp. \label{eq:pde2}
\end{align}

\section{Wave speed analysis}

Numerical simulation of \eqref{eq:origpde1}-\eqref{eq:origpde2} has shown that it exhibits travelling wave solutions, and it is our aim to determine their velocity  $c$. We will apply the same technique as for the Fisher equation \eqref{eq:fisher}, which involves characterising the fixed points of the corresponding autonomous system \cite{Gerlee2012,Murray}. 


\subsection{Transformation into autonomous system}
We will apply a similar kind of reasoning, and start by making the standard travelling wave ansatz $z=x-ct$ and move from PDEs to ODEs
\begin{align} \label{eq:ode} \nonumber
&cP'+f(P,M)=0\\
&cM'+ D_{\nu} ((1-P)M''+MP'') + g(P,M)=0 \nonumber 
\end{align}
where 
\begin{equation}\label{eq:fpm}
f(P,M)=\alpha(1-P-M) -(q_m+\mu)P + q_pM
\end{equation}
and
\begin{equation*}
g(P,M)=q_mP-(q_p+\mu)M
\end{equation*}
and prime denotes differentiation with respect to the new variable $z$. In order to proceed we want to transform the above ODEs to an autonomous system, and we do this by introducing new variables $Q=P'$ and $N=M'$. Now $Q=-f(P,M)/c$ and hence
\begin{equation*}
Q' = \frac{dQ}{dz} = -\frac{1}{c}\left( \frac{\partial f}{ \partial P}\frac{dP}{dz} + \frac{\partial f}{ \partial M}\frac{dM}{dz} \right) = -\frac{1}{c}\left( \frac{\partial f}{ \partial P}Q + \frac{\partial f}{ \partial M }N \right).
\end{equation*}
We now have the following autonomous system
\begin{align} \label{eq:notred}\nonumber
P'=&Q,  \\
M'=&N, \nonumber \\
Q'=& -\frac{1}{c}\left( \frac{\partial f}{ \partial P}Q + \frac{\partial f}{ \partial M }N \right), \nonumber \\
N' =& \frac{1}{ D_{\nu}(1-P)}\left( -cN- D_{\nu}MQ'-g(P,M) \right). \nonumber  \\
\end{align}
{We have previously analysed this system of equations numerically in order to calculate the wave speed \cite{Gerlee2012}. Below we show how the numerical approach can be avoided by reducing the dimensionality of the system, and then making use of the fact that $D_\nu$ is a large parameter to obtain an analytical estimate of the wave speed.}

Since \eqref{eq:fpm} is invertible we can use the relation $Q=-f(P,M)/c$ to express $P$ in terms of $Q$ and $M$, and hence reduce the dimensionality of the system. We now have
\begin{equation*}
-cQ=f(P,M)=\alpha P(1-P-M)-(q_m+\mu)P+q_pM \nonumber \\
\end{equation*}
which implies that
\begin{equation*}
\alpha P^2+(\alpha M-\alpha +q_m+\mu)P-q_pM-cQ=0. \nonumber \\
\end{equation*}
Since we will be interested in the dynamics near the origin we linearise the above equation and obtain
\begin{equation}\label{eq:P}
P=P^\star = \frac{q_pM+cQ}{q_m+\mu-\alpha}.
\end{equation}
Carrying out the linearisation of $f$ in equation \eqref{eq:notred} and inserting the above expression for $P$ results in the following three-dimensional system:
\begin{align} \label{eq:red} \nonumber 
M'=&N,  \\
\nonumber Q'=& -\frac{1}{c}\left( (\alpha-q_m-\mu)Q + q_pN \right),  \\
\nonumber N' =& \frac{q_m+\mu-\alpha}{(q_m+\mu-\alpha-q_pM-cQ)} \left( -cN +\frac{ D_{\nu}M}{c}\left( (\alpha - q_m-\mu)Q+q_pN\right), \right. \\
\nonumber & \left. + (q_p+\mu)M-q_m\frac{q_pM+cQ}{q_m+\mu - \alpha} \right).  \\
\end{align}
Here we can think of $P$ as being a fast variable in the system that quickly relaxes to a critical manifold defined by \eqref{eq:P}, and that the dynamics on this manifold is given by \eqref{eq:red}. In order to simplify further analysis of the system we will treat the special case $\mu = 0$. This is biologically motivated since in an untreated tumour the death rate of the cancer cells is generally much smaller than the proliferation rate, and hence $\mu \ll \alpha$.

\subsection{Phase space analysis}
The autonomous system \eqref{eq:red} has two fixed points, namely the trivial steady state $(M,Q,N)=(0,0,0)$ and the invaded state given by $(M,Q,N)=(q_m/(q_m+q_p),0,0)$. A travelling wave solution of the PDE-system \eqref{eq:pde1}-\eqref{eq:pde2}, corresponds to a heteroclinic orbit in the state space of the autonomous system, which connects the two steady states. In order to find the velocity of the traveling wave we will use a heuristic argument, which relies on non-negativity of $M(z)$, and hence on the characteristics of the fixed point at the origin. 

The orbit, which travels from the invaded fixed point to the fixed point at the origin, will only remain positive in $M$ if the fixed point at the origin 
is not a spiral. Precisely as with the Fisher equation, this depends on the wave speed $c$, and only for certain values of $c$ do non-negative orbits exists. We are looking for the smallest such value, which corresponds to the minimal wave speed of the system. 

A spiral at the origin is absent only if the eigenvalues of the Jacobian of the system \eqref{eq:red} are all real. The Jacobian of \eqref{eq:red} evaluated at the origin is given by
\begin{equation*}\label{eq:jac}
J(\mathbf{0}) = \left( \begin{array}{cccc}
0 & 0 & 1\\
0 & (q_m-\alpha)/c & -q_p/c\\
{D_{\nu}}^{-1} \left(q_p-q_mq_p/(q_m-\alpha)\right)& -q_mc/ (D_{\nu}(q_m-\alpha)) & -c/ D_{\nu} 
\end{array} \right).
\end{equation*}
The eigenvalues of $J$ are given by the roots of the characteristic equation
\begin{equation} \label{eq:char} 
P(\lambda) = \lambda^3 - (\frac{q_m-\alpha}{c} - c/D_\nu)\lambda^2 -D^{-1}_\nu(q_m+q_p-\alpha) \lambda -\frac{\alpha q_p}{D_\nu c},  
\end{equation}
and the smallest $c$ such that all roots of $P(\lambda)$ are real corresponds to the minimal wave speed. 

\subsection{Analysing the characteristic equation}
In order to find this $c$ we study the determinant of the polynomial, which for a general cubic equation $ax^3 + bx^2 + cx + d = 0$ is given by $\Delta = 18abcd - 4b^3d+b^2c^2 - 4ac^3 - 27a^2d^2$. Now if $\Delta < 0$ the equation has one real root and two complex roots, if $\Delta = 0$ then the equation has one multiple root and all roots are real, and if $\Delta > 0$ the equation has three distinct real roots. We are interested in the middle case $\Delta = 0$, which occurs precisely when the eigenvalues are all real.

The determinant of \eqref{eq:char} is however again a polynomial of degree three (but now in $c^2$), and in order to make progress we will make use of the fact that $D_\nu$ is a large parameter, and disregard terms of order $1/D^4_\nu$ and higher (and hence loose the $c^6$ term). This yields an approximation of the discriminant
\begin{eqnarray}\label{eq:dett}
 \hat{\Delta} (c) & = &-4A^3\alpha q_p\frac{1}{c^4D_\nu} + \left(A^2E^2 -  18AE\alpha q_p -27 \alpha^2 q_p^2 + 12 A^2 \alpha q_p \right)\frac{1}{c^2D^2_\nu} + \\ 
   && (4E^3 + 18E\alpha q_p -12A\alpha q_p - 2 AE^2)\frac{1}{D^3_\nu} \nonumber
\end{eqnarray}
where $A= q_m - \alpha$ and $E=q_m+q_p-\alpha$. 

We are looking for the smallest $c>0$, such that $\hat{\Delta}(c)=0$. Now $\hat{\Delta}(c)=0$ if and only if 
\begin{eqnarray*}\label{eq:det}
 \widetilde{\Delta} (c) & = &-4A^3\alpha q_p\frac{1}{D_\nu} + \left(A^2E^2 -  18AE\alpha q_p -27 \alpha^2 q_p^2 + 12 A^2 \alpha q_p \right)\frac{c^2}{D^2_\nu} + \\
   && (4E^3 + 18E\alpha q_p -12A\alpha q_p - 2 AE^2)\frac{c^4}{D^3_\nu}.
\end{eqnarray*}

\begin{figure}[!hptb]
\begin{center}
\includegraphics[width=15cm]{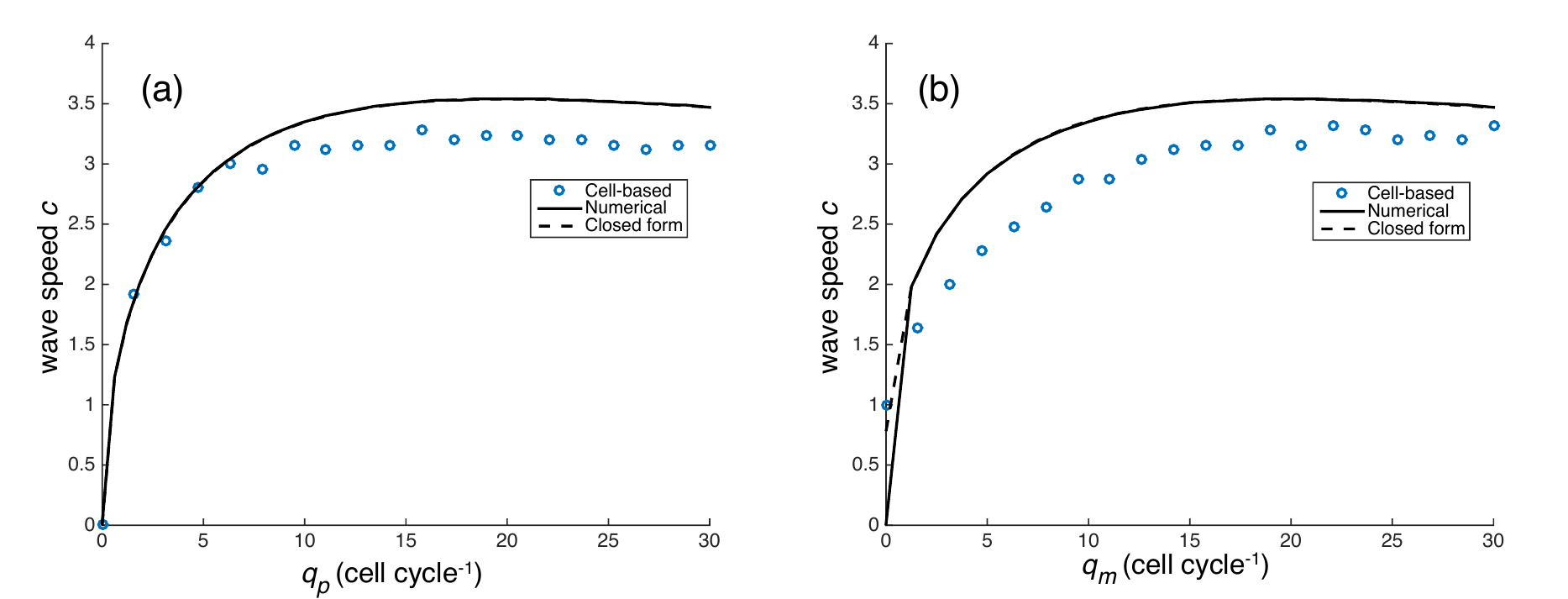}
\caption{\label{fig:comp}Comparison between wave speed calculated from the cell-based model, the Jacobian \eqref{eq:red}, and the analytical expression \eqref{eq:csol} of wave speed, when the parameters $q_m$ and $q_p$ are varied. In (a) $q_m=20$ and in (b) $q_p=20$. The other parameters are set to $\nu = 25$ and $\alpha = 1$.} 
\end{center}
\end{figure}

\begin{figure}[!hptb]
\begin{center}
\includegraphics[width=15cm]{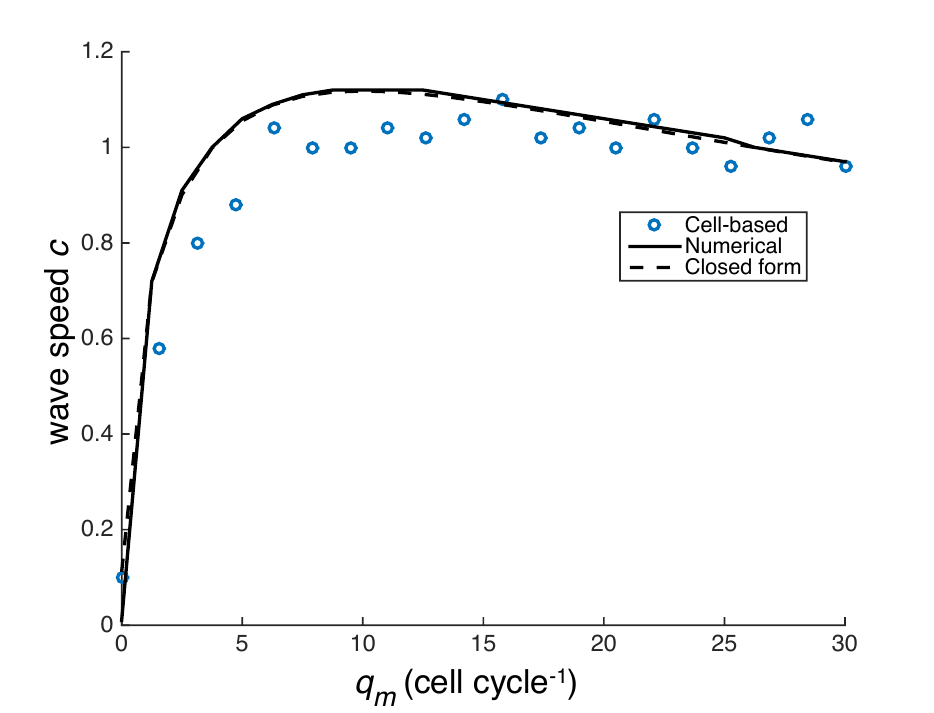}
\caption{\label{fig:compl}Comparison between wave speed calculated from the cell-based model, the Jacobian \eqref{eq:red}, and the analytical expression \eqref{eq:csol} of wave speed, when the proliferation is shifted $\alpha \rightarrow \alpha/10$. The other parameters are set to $q_p = 10$ and $\nu = 25$. Compared to fig. \ref{fig:comp}b the agreement is better, which is due to the larger difference between $\alpha$ and $D_\nu$.} 
\end{center}
\end{figure}

It is easily seen that the following statements about $\widetilde{\Delta}(c)$ hold:
\begin{enumerate}
\item $\widetilde{\Delta}(0) = -4A^3\alpha q_p < 0$
\item $\widetilde{\Delta}'(c) > 0$ for $c>0$
\item $\lim_{c \rightarrow \infty} \widetilde{\Delta}(c) = \infty$
\end{enumerate}
From the above statements, and since $\widetilde{\Delta}(c)$ is a quadratic in $s=c^2$, we know that there exists only one $c_m \in (0,\infty)$ such that $\widetilde{\Delta}(c_m) = 0$. Since the zeros of $\widetilde{\Delta}(c)$ and $\Delta(c)$ coincide we know that also $\Delta(c_m) = 0$ holds. In order to find this minimal $c$ we carry out the variable substitution $s=c^2$, and solve the resulting quadratic to get
\begin{eqnarray*} \label{eq:ssol}
s_m= \pm D_\nu \frac{ \sqrt{K^2 + I} - K}{J}
\end{eqnarray*}
where
\begin{eqnarray*} 
K&=& E^2A^2-18AE \alpha q_p-27 \alpha^2q_p^2 + 12 A^2 \alpha q_p,\\
I &=& 16A^3\alpha q_p (4E^3 + 18E\alpha q_p -12A\alpha q_p - 2 AE^2),\\
J &=& 2(4E^3 + 18E\alpha q_p -12A\alpha q_p - 2 AE^2).
\end{eqnarray*}
The minimal wave speed is now given by $c_m=\pm \sqrt{s_m}$, and since the velocity is positive and real we can disregard the negative and complex solution, and hence get
\begin{align} \label{eq:csol}
c_m= \sqrt{ \frac{D_\nu}{J} \left( \sqrt{K^2 +I} - K \right)}.
\end{align}
We can now compare this expression with the wave speed obtained by simulating the cell-based model and by numerically calculating the eigenvalues of the Jacobian \eqref{eq:red}. This comparison shows that the closed form expression gives a good approximation of the propagation speed of the invading cancer cells when $q_m$ is large, but overestimates the speed for low $q_m$ (fig.\ \ref{fig:comp}). The reason for this is that the derivation of the continuum description requires the assumption that migration occurs much more frequently than proliferation. For the stochastic process that underlies the Fisher equation this implies $\alpha \ll D_\nu$, but for our model where migration only occurs in one phenotypic state the relative values of $q_p$ and $q_m$ also matter. It is however possible to improve the agreement between the derived wave speed and the speed of propagation in the individual-based model by reducing $\alpha$, as can be seen in fig.\ \ref{fig:compl} where $\alpha$ has been reduced by a factor ten. It is also worth noting that the discrepancy between the closed form solution and the wave speed obtained by analysing the Jacobian is minimal, suggesting that the simplification of the discriminant was justified. 

{It is also possible to consider an even stronger simplification by ignoring terms of order $1/D_{\nu}^3$ and higher in the determinant \eqref{eq:dett}. Although this simplification yields a similar $\sqrt{D_{\nu}}$ scaling in the velocity, the numerical values of the wave speed are incorrect in this case and deviate by more than a factor of three.}

{Lastly we note that it is known that continuous descriptions, in terms of PDEs, of discrete systems typically tend to overestimate the wave speed of invading fronts \cite{Brunet1997}. This source of error in part explains the overestimate of the wave speed that is seen in figure \ref{fig:comp} and \ref{fig:compl}.}

\subsection{Limiting wave speed $c^\ast$}
The above derived expression for $c$ is complicated, and it seems difficult to draw any conclusions about the impact of the model parameters by simply inspecting the formula. One exception is the diffusion constant $D_\nu$, which only appears once in the expression, and it is clear that $c \sim \sqrt{D_\nu}$, just as for the Fisher equation. 

\begin{figure}[!hptb]
\begin{center}
\includegraphics[width=15cm]{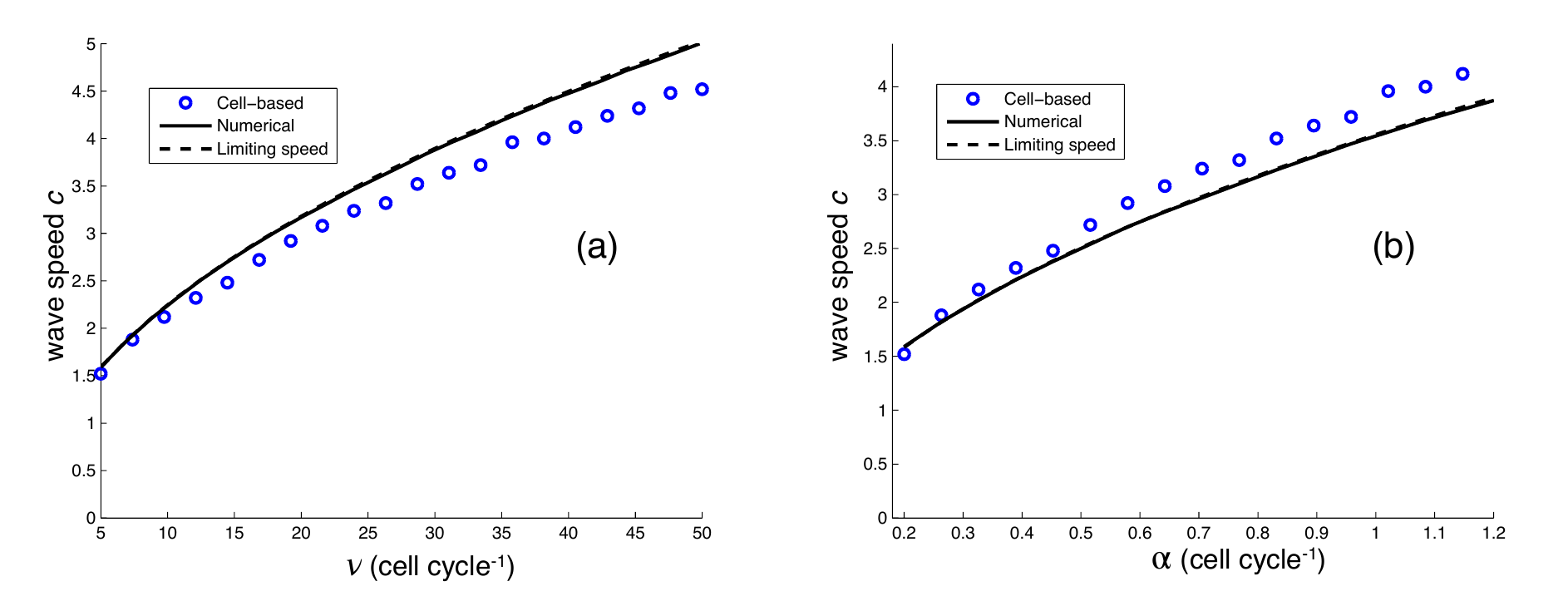}
\caption{\label{fig:climit}Comparison between wave speed of the cell-based model, obtained from the Jacobian \eqref{eq:red} and the limiting expression \eqref{eq:climit}, valid when the switching rates satisfy $q_p=q_m \gg \alpha$. In (a) migration rate $\nu$ is varied and in (b) the proliferation rate $\alpha$ is varied. The switching rates are set to $q_p=q_m=30 \gg \alpha = 1$.} 
\end{center}
\end{figure}

In order to gain further insight into how the parameters influence the wave speed we will consider the case when $q_m=q_p=q$, and $\alpha \ll q$. Biologically this means that the phenotypic switching rates to and from the proliferative and migratory states are equal, and much larger than the proliferation rate of the cells (i.e.\ switching typically occurs many times between two cell division events). We proceed by expanding the expressions for $I$, $J$ and $K$, and since $\alpha \ll q$, we only retain zeroth and first order terms in $\alpha$. This yields:

\begin{eqnarray*} 
K&=&(2q-\alpha)^2(q-\alpha)^2-18(2q-\alpha)(q-\alpha) \alpha q-27 \alpha^2q^2 + 12\alpha q (q-\alpha)^2 \approx 4q^4-36\alpha q^3,\\
I &=& 16(q-\alpha)^3\alpha q \left( 4 (2q-\alpha)^3 -2(q-\alpha)(2q-\alpha)^2 \right) \approx 384 \alpha q^7, \\
J &=& 2 \left( 4(2q-\alpha)^3 + 18 (2q-\alpha)\alpha q -12 (q-\alpha)\alpha q \right) \approx 48 q^3 - 16 \alpha q^2.
\end{eqnarray*}
Now $\sqrt{K^2 + I} \approx 4q^4\sqrt{1+6\alpha / q}$, and we proceed by a first order Taylor expansion of the square root term ($\sqrt{1+x} \approx 1 + 1/2x$, in the variable $x=6\alpha / q$)  to obtain $\sqrt{K^2 + I} \approx 4q^4+12\alpha q^3$. Finally we can write $\sqrt{K^2 + I} - K \approx 48 \alpha q^3$, and by rearranging the terms we arrive at the expression

\begin{equation}\label{eq:climit}
c^\ast=\sqrt{\frac{D_v \alpha}{1-\alpha/3q}}.
\end{equation}
In the limit $q\rightarrow \infty$ this reduces to $c^\ast=\sqrt{D_\nu \alpha}$, which is precisely half of the wave speed of the Fisher equation. In fact, this is hardly surprising, since in the microscopic view, the cells are spending half the time in an immobile proliferative state, which reduces their total mobility by one half. A comparison between the wave speed $c^\ast$, and the actual wave speed of the system is shown in fig. \ref{fig:climit}.

\section{Asymptotic solution}

In the previous section we established a relation between the model parameters and the speed at which the tumour grows, but it would also be useful to know how it grows, i.e. how the parameters affect the shape of the invading front. We therefore proceed with a derivation of an approximate solution to our system \eqref{eq:pde1} - \eqref{eq:pde2}. Again we consider the case $\mu = 0$, which, as noted above, is biologically plausible. This implies that we again are dealing with the following system of coupled ODEs:
\begin{align} \label{eq:ode2}
&cP'+\alpha P(1-P-M) -q_m P + q_pM=0\\
&cM'+D_\nu((1-P)M''+MP'')  - q_pM + q_m P=0 \nonumber 
\end{align}
but now with the aim of finding approximate solutions  $P(z)$ and $M(z)$. We start by noting that among the coefficients of the above ODEs, the proliferation rate $\alpha$ is smaller than the other parameters, 
and motivated by this we will attempt to find a solution using a standard singular perturbation technique, and express the solution as a Taylor expansion in the parameter $\alpha$. 

We start by fixing the solution along the $z$-direction, such that $P(z) + M(z)=1/2$ at $z=0$, and introduce a change in variables $\xi = \alpha z$, and look for solutions $f(\xi)=P(z)$ and $g(\xi) = M(z)$. 

\begin{figure}[!hptb]
\begin{center}
\includegraphics[width=13cm]{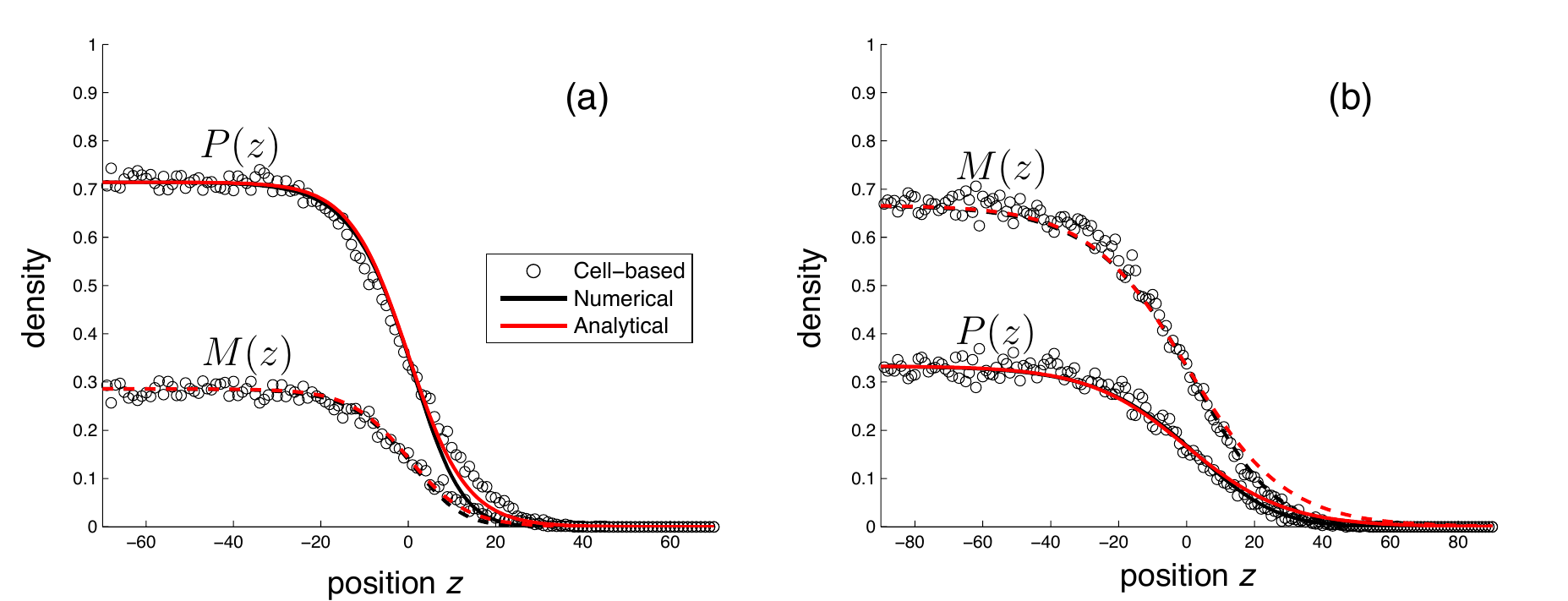}
\caption{\label{fig:frontcomp}Comparison between numerical solutions of the full system (1)-(2), and the analytical solutions \eqref{eq:leading} obtained through a singular perturbation approach. Solid lines show $P(z)$ and dashed lines $M(z)$. In (a) $(q_p,q_m)=(25,10)$, while in (b) we have $(q_p,q_m)=(10,20)$. The other parameters are fixed at $\nu = 25$ and $\alpha = 1$. The initial condition for the numerical solutions was given by $p(x,0)=\exp(-10x)$ and $m(x,0)=0$.} 
\end{center}
\end{figure}

This change of variables transforms \eqref{eq:ode2} to
\begin{align} \label{eq:ode3}
&\alpha c f'+\alpha f(1-f-g) -q_m f + q_p g =0\\
&\alpha c g'+\alpha^2 D_\nu((1-f)g''+gf'')  - q_pg + q_m f=0. \nonumber 
\end{align}
with boundary conditions
\begin{align} \label{eq:bc} \nonumber
& \lim_{\xi \rightarrow \infty} f(\xi)+g(\xi)=0, \\
& \lim_{\xi \rightarrow -\infty} f(\xi)+g(\xi)=1,\\ \nonumber
&  f(0)+g(0)=1/2. \nonumber
\end{align}
We now look for solutions of the form
\begin{align} \label{eq:ans}
& f(\xi) = f_0(\xi) + \alpha f_1(\xi) + \mbox{H.O.T.}\\
& g(\xi) = g_0(\xi) + \alpha g_1(\xi) + \mbox{H.O.T.} \nonumber
\end{align}
Since this should be valid for all values of $\alpha$, the above boundary conditions \eqref{eq:bc} transform to
\begin{align} \nonumber
& f_0(-\infty) = g_0(-\infty)= 1 & f_1(-\infty) = g_1(-\infty)= 0\\ \nonumber
& f_0(\infty)=g_0(\infty)=0 & f_1(\infty)=g_1(\infty)=0\\ \nonumber
& f_0(0)+g_0(0)=1/2 & f_1(0)+g_1(0)=0. \nonumber
\end{align}
On substituting \eqref{eq:ans} into \eqref{eq:ode3} and equating powers of $\alpha$ we get
\begin{align} 
& O(1): q_mf_0-q_pg_0 = 0 \label{eq:order11}\\
& O(\alpha): cf_0' + f_0 (1-f_0-g_0)-q_mf_1+q_pg_1=0 \label{eq:order12}
\end{align}
and
\begin{align} 
& O(1): q_mf_0-q_pg_0 = 0 \label{eq:order21}\\
& O(\alpha): cg_0'+q_mf_1-q_qg_1 = 0. \label{eq:order22}
\end{align}
By combining \eqref{eq:order12} with \eqref{eq:order11} and \eqref{eq:order22} we obtain the following equation for $f_0$:
\begin{equation*}\label{eq:f0}
c(1+\rho)f_0'+f_0(1-(1+\rho)f_0)=0
\end{equation*}
with boundary condition $f_0(0) + g_0(0)=1/2$, or equivalently $f_0(0)=1/2(1+\rho)$, where $\rho=q_m/q_p$. A solution to this equation is given by
\begin{equation*}
f_0(\xi) = \frac{1}{(1+\rho)(1+e^{\frac{\xi}{(1+\rho)c}})}
\end{equation*}
and from \eqref{eq:order11} we can now calculate $g_0$ as
\begin{equation*}
g_0(\xi) = \frac{\rho}{(1+\rho)(1+e^{\frac{\xi}{(1+\rho)c}})}.
\end{equation*}
In terms of the original variable $z$ we now get the following approximate solutions:
\begin{align}\label{eq:leading}
&P(z) = \frac{1}{(1+\rho)(1+e^{\frac{\alpha z}{(1+\rho)c}})},\\
&M(z) = \frac{\rho}{(1+\rho)(1+e^{\frac{\alpha z}{(1+\rho)c}})}. \nonumber
\end{align}
One could of course try to obtain higher order solutions, but the equations encountered are non-linear and do not permit closed-form solution, and the leading order solution \eqref{eq:leading} is in fact very close to numerical solutions of the system (see fig. \ref{fig:frontcomp}).

These analytical solutions makes it possible to relate the steepness $s$ of the solution to the parameters of the model. A reasonable measure of the steepness is the slope of the total density of glioma cells at $z=0$, which is given by
\begin{equation}
s = -\frac{d}{dz}(P(z)+M(z))\Big|_{z=0} = \frac{\alpha}{4c(1+\rho)}.
\end{equation}
This can be compared with the corresponding quantity for the Fisher-equation, which is given by $s_{FE}=1/4c$, and shows that the switching dynamics, represented by the $(1+\rho)$-term in the denominator, makes the front less steep, or in other words, the tumour margin more diffuse. Further, the result for the Fisher equation stating that a faster front always is less step does no longer hold, since one can construct a solution with a small wave speed $c$, but with large a $\rho$. 

\section{Discussion and conclusion}
In this paper we have analysed the behaviour of a cell-based model of glioblastoma growth, via the analysis of its continuum approximation, and focused on the properties of travelling wave solutions. In particular we have derived an approximate closed form expression for the wave speed, and a leading order approximation of the shape of the invading front. Agreement is good when cell migration dominates over proliferation, which is to be expected since this assumption underlies the derivation of the PDE description. 

The expression for the wave speed we have derived correspond to the minimal speed, and it is currently now know for which initial conditions this wave speed is attained. The results from the cell-based model do however suggest that the average cell density profile in the stochastic setting advances at the minimal wave speed. 


This system of equations does not only describe tumour growth, but applies to any spatially extended population in which the individuals switch between a motile and stationary/proliferative state. In fact a similar system, that lacks a non-linearity in the diffusion term, has been analysed by Lewis and Schmitz \cite{Lewis1996}. We have followed their approach, but where they rely on numerical solutions to the eigenvalue problem we have shown that an approximate, but highly accurate, solution can be obtained by analysing a truncated version of the determinant. In addition they derived an asymptotic solution only valid when the switching rates are equal, whereas we have considered the general case $q_m\neq q_p$. The analysis presented here can therefore be seen as a natural extension to their work.

When comparing our results to more recent work in glioblastoma modelling it is worth noting that many studies have shown the square root scaling of the wave speed as a function of diffusion coefficient (or migration rate) and proliferation rate \cite{Gerlee2012,Hatzikirou2009,Fedotov2008}, but this is the first study to report a closed form expression of the wave speed, which explicitly involves the switching rates $q_{p,m}$. The derivation of the front shape as a function of $q_{p,m}$ is also a new result, which brings into light an interesting difference between the two-component system and the Fisher equation. For the latter, the steepness of the front is given by $1/4c$, which implies that faster invasion corresponds to a less steep front. We have shown that when the cells switch between proliferation and migration,  the steepness is equal to $\alpha/4c(1+\rho)$, and since one can construct solutions with a small velocity $c$, but with a large $\rho$, the result for the Fisher equation does no longer hold. In other words one can have fronts with a small slope that still move slowly. 


{Lastly, we note that our results are encouraging for those who hope to bridge different scales in tumour biology using mathematical analysis. Our results suggests that it is possible to connect the microscopic properties of cells to the macroscopic outcomes of tumour growth. Although this study is theoretical we believe that it will increase our understanding of this inherently multi-scale disease.}

\section{Acknowledgments}
The authors would like to thank Bernt Wennberg for pointing us in the right direction during the initial phases of analysis, and Assar Gabrielsson Fond, Cancerfonden, Swedish Research Council (grant no.\ 2014-6095) and Swedish Foundation for Strategic Research (grant no.\ AM13-0046) for funding.

\bibliography{glioma}
\bibliographystyle{elsarticle-num}







\end{document}